# Photonic Spin Hall Effect in Waveguides Composed of Two Types of Single-Negative Metamaterials

**Zhiwei Guo[1], Haitao Jiang[1,*], Yang Long[2], Kun Yu[1], Jie Ren[2], Chunhua Xue[3] & Hong Chen[1]**

[1] Key Laboratory of Advanced Micro-structure Materials, MOE, School of Physics Science and Engineering, Tongji University, Shanghai 200092, China.
[2] Center for Phononics and Thermal Energy Science, China-EU Joint Center for Nanophononics, Shanghai Key Laboratory of Special Artificial Microstructure Materials and Technology, School of Physics Sciences and Engineering, Tongji University, 200092 Shanghai, China
[3] School of Computer Science & Communication Engineering, Guangxi University of Science and Technology, Liuzhou, Guangxi 545006, China
* Correspondence and requests for materials should be addressed to H.T.J. (email: jiang-haitao@tongji.edu.cn)

## ABSTRACT

The polarization controlled optical signal routing has many important applications in photonics such as polarization beam splitter. By using two-dimensional transmission lines with lumped elements, we experimentally demonstrate the selective excitation of guided modes in waveguides composed of two kinds of single-negative metamaterials. A localized, circularly polarized emitter placed near the interface of the two kinds of single-negative metamaterials only couples with one guided mode with a specific propagating direction determined by the polarization handedness of the source. Moreover, this optical spin-orbit locking phenomenon, also called the photonic spin Hall effect, is robust against interface fluctuations, which may be very useful in the manipulation of electromagnetic signals.

## Introduction

The spin Hall effect (SHE) adds a plenty of physics for the family of Hall effect in solid-state physics. Different from the usual Hall effect, in which external magnetic fields are needed to break the time-reversal ($T$) symmetry of the system, the SHE can occur in $T$-symmetric electron systems with spin-orbit interaction[1]. Because of the spin-orbit or spin-momentum locking, electrons with opposite spin will take different trajectories. In analogy to the SHE of electrons, in some cases photons with different circular polarizations (optical spin) may go different ways, which can be called photonic spin Hall effect (PSHE). This optical spin-orbit locking phenomenon possesses many important applications in photonics and has attracted the attentions of more and more people in recent years (see Refs. [2-8] and the references therein). In particular, in 2015 Bliokh *et al.* theoretically reveal that the optical spin-orbit locking and thereby the PSHE can occur for the surface guided modes, owing to the transverse spin property of evanescent waves[9]. Their remarkable findings discover the unusual spin of evanescent waves and deepen our understanding of the Maxwell's theory. The theory in Ref. [9] can explain a variety of PSHEs connected with localized optical modes such as the surface plasmon polaritons (SPPs) at air-metal interfaces[10-12], the guided modes in the nanofiber[13,14] and the subwavelength modes in hyperbolic metamaterials[15]. For the surface guided modes, the property of evanescent waves plays a crucial role in the formation of PSHE. Up to now, people mainly study the evanescent waves in the $\varepsilon$-negative (ENG) material in which the permittivity is negative, as in the case of SPPs. However, another kind of single-negative (SNG) metamaterial, that is the $\mu$-negative (MNG) metamaterial in which the permeability is negative, also supports the evanescent waves[16,17]. In addition, it has recently been shown that ENG and MNG metamaterials possess different topological properties from the viewpoints of band-edge inversion[18,19] and Berry potential[20,21], respectively. Moreover, Silveirinha theoretically reveals a PSHE in a kite-shaped interface composed of continuous ENG and MNG metamaterials[21]. Since this theoretical work offers important applications for spin-directional optical interfaces, the related experiments are highly desirable. Although the metal is a natural ENG material below plasma frequency, it is still a great challenge to fabricate a MNG material. Nevertheless, in microwave regime, the two-dimensional (2D) transmission lines (TLs) loaded with lumped circuit elements provide a convenient platform to realize various effective $\varepsilon$ and $\mu$ and observe unusual wave propagations[22,23]. Therefore, it would be possible to experimentally observe the PSHE in a waveguide composed of ENG and MNG metamaterials (for convenient, we call it ENG/MNG waveguide in the following parts) based on the 2D TLs.

In this paper, by loading lumped inductors and capacitors into 2D TLs, we fabricate effective ENG and MNG metamaterials, respectively, and study the guided modes in ENG/MNG waveguide. By putting a localized emitter near the ENG/MNG interface, we find that the propagating direction of the excited guided mode is determined by the polarization handedness of the source. For a linearly polarized source, it will couple equally with two guided modes oppositely propagating along the interface. However, for a circularly polarized source with a specific spin direction, it will excite one guided mode with a specific propagating direction, which demonstrate the PSHE. Moreover, we find that the PSHE is robust against interface fluctuations, which may be very useful in the manipulation of electromagnetic signals.

## Results

**Guided modes in ENG/MNG waveguides based on 2D TLs.** The schematic of PSHE in an ideal ENG/MNG waveguide is shown in Fig. 1. A circularly polarized source is put near the ENG/MNG interface. The source with specific handedness only excites one guided mode with a specific propagating direction. This optical spin-orbit locking phenomenon gives PSHE. Now we construct a real ENG/MNG waveguide based on 2D TLs. TLs with lumped elements can realize various effective electromagnetic parameters[24,25]. Our 2D TLs are fabricated on a FR-4 substrate with a thickness of $h = 1.6$ mm and relative permittivity of $\varepsilon_r = 4.75$. The width of the TLs is $w = 1$ mm and the length of the unit cell is $d = 10$ mm. Structural factor of the TL is defined as $g = Z_0 / \eta_{eff}$, where $Z_0$ and $\eta_{eff}$ are the characteristic impedance and the effective wave impedance of the normal TL, respectively. When $w < h$, $g = \frac{1}{2\pi} \ln(\frac{8h}{w} + \frac{w}{4h}) = 0.408$. For the normal TL, $\varepsilon_{normal-TL} = \frac{\varepsilon_r + 1}{2} + \frac{\varepsilon_r - 1}{2} \cdot \frac{1}{\sqrt{1 + 12h/w}}$ and $\mu_{normal-TL} = 1$. Then, the effective permittivity and permeability of 2D TLs with lumped elements can be written as[22,23]:

$$\begin{aligned}\varepsilon &= (2C_0 \cdot g - \frac{g}{\omega^2 \cdot L \cdot d})/\varepsilon_0 \\ \mu &= (\frac{L_0}{g} - \frac{1}{\omega^2 \cdot C \cdot d \cdot g})/\mu_0\end{aligned}, \qquad (1)$$

where $\omega$ denotes the angular frequency, $L$ and $C$ denote the shunted lumped inductors and the series lumped capacitors, respectively. In Eq. (1),

$$\begin{aligned}C_0 &= \varepsilon_{normal-TL} \cdot \varepsilon_0 / g \\ L_0 &= \mu_{normal-TL} \cdot \mu_0 \cdot g\end{aligned}, \qquad (2)$$

where $\varepsilon_0$ and $\mu_0$ are the permittivity and permeability of vacuum, respectively. From Eqs. (1) and (2), the effective permittivity and permeability of 2D TLs can be calculated. Tuning the values of loaded elements $L$ and $C$, one can realize an effective ENG or MNG metamaterials. In our paper, we consider a simple situation in which the effective ENG metamaterial is realized only by loading series lumped capacitors $C=0.5$ pF and the effective MNG metamaterials is realized only by loading shunted lumped inductors $L=1$ nH into the TLs. In this case, the effective parameters are shown in Fig. 2(a). ENG or MNG metamaterials can be realized once the frequencies are lower than the cutoff frequency. In Fig. 2(a), the spectrum is divided into three parts and the waveguide composed of two materials is schematically shown in the inset. Permittivity and permeability of the two materials are shown by the dashed and solid lines, respectively. ENG-MNG region painted by the light blue corresponds to the frequency range below 3.14 GHz. Above this cutoff frequency, the parameters of two materials become ENG and double-positive (DPS), respectively, in the frequency range from 3.14 GHz to 4.2 GHz. This is the ENG-DPS region painted by the light orange. Above the frequency of 4.2 GHz, the white region denotes the DPS-DPS region.

Based on the parameters in Fig. 2(a), we calculate the dispersion relation of guided modes in ENG/MNG waveguide for different polarizations of the incident wave. Since the tangential components of electric field and magnetic field should be continuous at the boundary of ENG and MNG materials, the dispersion relation of surface guided modes for transverse electric (TE) polarization can be deduced by the characteristic equation $k_{y1}/k_{y2}=-\mu_1/\mu_2$[26,27]. While for the transverse magnetic (TM) polarization it is $k_{y1}/k_{y2}=-\varepsilon_1/\varepsilon_2$. In the characteristic equation, $k_{yi}=\sqrt{k_x^2-k_i^2}$, where $i=1,2$ correspond to two different materials, $k_x$ and $k_y$ denote the propagation constants of the guided mode along and perpendicular to the interface, respectively. After some deductions, we get the following dispersion equations:

$$k_x = \pm \frac{\omega}{c} [\frac{\mu_1 \mu_2}{\mu_1^2 - \mu_2^2}(\mu_1 \varepsilon_2 - \varepsilon_1 \mu_2)]^{1/2} \quad (TE\ polarization)$$
$$k_x = \pm \frac{\omega}{c} [\frac{\varepsilon_1 \varepsilon_2}{\varepsilon_2^2 - \varepsilon_1^2}(\mu_1 \varepsilon_2 - \varepsilon_1 \mu_2)]^{1/2} \quad (TM\ polarization)$$

, (3)

where $c$ is the speed of light in vacuum. $\varepsilon_i$ and $\mu_i (i=1,2)$ are the parameters of two different materials. The corresponding dispersion relations based on Eq. (3) are shown in Fig. 2 (b). It is interesting that the slopes of dispersions for TE and TM polarizations are different. For TM (TE) polarization, the slope of dispersion is positive (negative) and the guided modes are forward (backward) surface waves for which the direction of wave-vector component is parallel (anti-parallel) with that of Poynting-vector component along the interface. Because of the boundary conditions, the components of Poynting vectors (energy flows) along the interface on the two sides of the ENG/MNG interface are in opposite directions, see the similar analysis of TM waves in Fig. 3 in Ref. [28]. The direction of the group velocity along the interface is determined by the net Poynting-vector component after the compensation between the Poynting-vector components on the two sides of the interface. For TM (TE) waves, the direction of the net Poynting-vector component along the interface is from the left (right) to the right (left). So the TM (TE) waves are forward (backward) surface waves. Moreover, using suitable parameters, we can tune the net Poynting-vector component along the interface to be near zero and obtain the ultra-slow wave with a near-zero group velocity.

**Transverse spin in evanescent waves and the spin-orbit locking effect.** The transverse spin of evanescent wave plays an important role in the formation of PSHE. In this section, we use the designed parameters in Fig. 2(a) to systematically study the transverse spin of evanescent waves in ENG and MNG metamaterials, respectively. Just as schematically shown in Fig. 1, there is an interface composed by an ENG metamaterial in the upper part and a MNG metamaterial in the lower part at $y = 0$. Taking TM wave as an example, the generic evanescent wave solution of Maxwell equations in MNG and ENG metamaterials can be written as[3,9]:

$$\vec{E}_{MNG} = \begin{pmatrix} -i\dfrac{k_1}{\varepsilon_1 \omega} \\ \dfrac{k_x}{\varepsilon_1 \omega} \\ 0 \end{pmatrix} \cdot e^{ik_x x - k_1 y}, \quad \vec{E}_{ENG} = \begin{pmatrix} i\dfrac{k_2}{\varepsilon_2 \omega} \\ \dfrac{k_x}{\varepsilon_2 \omega} \\ 0 \end{pmatrix} \cdot e^{ik_x x + k_2 y}, \tag{4}$$

where $\vec{E}_{MNG}$ and $\vec{E}_{ENG}$ are the electric fields in MNG and ENG metamaterials, respectively. The corresponding "wavefunctions" can be written as[9]:

$$\psi_{Surf}(MNG) = \begin{pmatrix} -i\dfrac{k_1}{\varepsilon_1 \omega} \\ \dfrac{k_x}{\varepsilon_1 \omega} \\ 0 \\ 0 \\ 0 \\ 1 \end{pmatrix}, \quad \psi_{Surf}(ENG) = \begin{pmatrix} i\dfrac{k_2}{\varepsilon_2 \omega} \\ \dfrac{k_x}{\varepsilon_2 \omega} \\ 0 \\ 0 \\ 0 \\ 1 \end{pmatrix}. \tag{5}$$

The spin angular momentum of light in the $z$ direction is described by the operator $\hat{S}_z = -i\begin{pmatrix} 0 & 1 & 0 \\ -1 & 0 & 0 \\ 0 & 0 & 0 \end{pmatrix}$. Using the operator $\hat{\Sigma} = \begin{pmatrix} \hat{S} & 0 \\ 0 & \hat{S} \end{pmatrix}$, we can obtain the spin density in a plane wave in the $z$ direction[9]:

$$\vec{S}_{surf} = \psi_{surf}^{\dagger} \hat{\Sigma} \psi_{surf}. \tag{6}$$

Based on Eqs. (3), (5) and (6), we can calculate the transverse spin of evanescent waves for ENG and MNG metamaterials, which are indicated by the green and yellow balls, respectively, in Fig. 3. It is seen that, for the spin-up ($\overset{\scriptscriptstyle 1}{S}_{surf} > 0$), the wavevector or momentum in the MNG (ENG) metamaterial is positive (negative). However, for the spin-down ($\overset{\scriptscriptstyle 1}{S}_{surf} < 0$), the momentum in the MNG (ENG) metamaterial is negative (positive). Therefore, for the different spin in the same material (MNG or ENG metamaterial), the direction of corresponding momentum along the interface should be inverted, which demonstrates the spin-momentum locking effect. To further confirm this effect, we numerically calculate the magnetic-field distributions in the $z$ direction at frequency of 2.7 GHz for rotating source with different handedness located in ENG or MNG metamaterial by using COMSOL MULTI-PHYSICS, as are shown in the four insets in Fig. 3. At frequency of 2.7 GHz, the absolute value of $k_x$ is $2.13k_0$, which are marked by the red dashed lines in Fig. 3. From the magnetic-field distributions, one can see that, for the anti-clockwise rotating source [denote by $(1, i)$] corresponding to the spin-up ($\overset{\scriptscriptstyle 1}{S}_{surf} > 0$) in MNG (ENG) metamaterial, the direction of the guided mode along the interface is from the left (right) to the right (left). However, for the clockwise rotating source [denote by $(1, -i)$] corresponding to the spin-down ($\overset{\scriptscriptstyle 1}{S}_{surf} < 0$) in MNG (ENG) metamaterial, the direction of the guided mode along the interface is from the right (left) to the left (right). The calculated magnetic-field distributions clearly show the PSHE in ENG/MNG waveguide. It is worth noting that, although we only study the TM wave, the transverse spin of TE wave and thereby the PSHE can also be obtained by using the same method.

**Experimental demonstration of the PSHE in ENG/MNG waveguides.** In this section, we use the designed parameters in Fig. 2(a) to experimentally demonstrate the PSHE. In general, to observe the beam splitting effect, an excited source with a specific polarization is needed. For example, one can use a circularly polarized light through the Rayleigh scattering of small particle [4]. Another way is based on the rotating electric dipole ($\vec{P} = \vec{P}_x \pm i\vec{P}_y$) or magnetic dipole ($\vec{M} = \vec{M}_x \pm i\vec{M}_y$) near the interface for different polarizations[10]. TM (TE) guided modes are excited by an in-plane rotating electric (magnetic) dipole. Different from the above two methods, by using the 2D TLs we can also design a source with a specific polarization, which has been successfully used to observe the PSHE in hyperbolic metamaterial[15]. Considering the symmetry of the structure, we design two samples to perform the simulations and experiments.

The first sample contains an ENG metamaterial in the upper part and a MNG metamaterial in the lower part. The sample has $12 \times 13$ unit cells for the symmetry of structure, as is shown in Fig. 4(a). A linearly polarized source is loaded near the center of the sample marked by the red point. In simulations, a 1V voltage source is used to act as a linearly polarized source. At the edges of two samples, matching resistors with $R = 85\,\Omega$ are loaded to avoid the influence of reflected waves. Inset shown in the right is an amplified capacitors and inductor in the unit cell of ENG and MNG metamaterials, respectively. The corresponding circuit model is also beside it, where the interface is showed by the gray dashed line. For the 2D field distribution, we measure the vertical component of magnetic fields that are denoted by $|H_z|$. We use the CST (computer simulation technology) microwave studio software to perform the simulation. In Fig. 4(b) we simulate $|H_z|$ patterns of the sample when a TM-polarized guided mode at 2.7 GHz as shown in Fig. 2(b) is excited. It is seen that a source with a linearly polarized polarization couples equally with the two guided modes which oppositely propagate along the interface. The measured $|H_z|$ patterns 1 mm above the surface of the sample at 2.7 GHz are shown in Fig. 4(c). It is seen that the measured $|H_z|$ patterns in Fig. 4(c) agree well with the simulated results in Fig. 4(b), both of which demonstrate that a linearly polarized source without spin property has no direction-selective excitation of guided modes.

However, the excitation of guided modes for the circularly polarized source with spin property is quite different. The prototype of the second sample used in simulations and experiments is composed of 12×12 unit cells, as is shown in Fig. 5(a). An amplified circuit model shows that there is a 90 degree phase difference between the four linearly polarized sources (marked by the red points) in the anticlockwise direction. These four linearly polarized sources as a whole can take effect of a circularly anticlockwise-rotating source. If the 90 degree phase difference between the four linearly polarized sources is in the clockwise direction, an effective circularly clockwise-rotating source will be realized. Therefore, by controlling the phase difference between the four sources, one can flexibly realize left- or right-handed rotating source just as done in Ref. [15]. Now we put two kinds of rotating sources with different handedness near the ENG/MNG interface and study the excitation of the guided modes, respectively. In Fig. 5(b), we both simulate and measure $|H_z|$ patterns excited by the rotating source. The parameters in Fig. 5 are the same as those in Fig. 4 except that the sample size and the source are different. Overall, the simulated $|H_z|$ patterns are in good agreement with the experimental results. For the clockwise-rotating source, only the guided mode propagating from right to left is excited, as shown in Fig. 5(b) and 5(c). However, for the anticlockwise-rotating source, only the guided mode propagating from left to right is excited, as shown in Fig. 5(d) and 5(e). In other words, the selective excitation of guided mode with specific propagating direction is determined by the polarization handedness of the source. The unidirectional wave transport originates from the optical spin-orbit locking. To quantitatively see the unidirectional transmission property, in Figs. 5(f) and 5(g) we perform the average $|H_z|$ in the interface of ENG and MNG metamaterials. In Fig. 5(f) the solid lines give the simulated values along the ENG/MNG interface while the measured average values in the same region are shown by the dashed lines in Fig. 5(g). The direction of unidirectional transmission will invert when the rotating direction of source reverses. So far we have experimentally demonstrated the PSHE in ENG/MNG waveguides.

Finally, since the interface fluctuation is an important factor in practice, here we study the influence of the interface fluctuation on the PSHE. We introduce some bulges in the previous flat interface, as is schematically shown in Fig. 6(a). Then we numerically study the PSHE in this imperfect ENG/MNG interface. The sample in Fig. 6(b) or Fig. 6(c) is the same as that in Fig. 5(b) or Fig. 5(d) except that the flat interface is changed to an imperfect interface with some bulges. From the simulated field patterns in Figs. 6(b) and 6(c), we see that a circularly clockwise-rotating (anticlockwise-rotating) only excite the guided mode propagating from right (left) to left (right). Therefore, the PSHE occurring in the ENG/MNG waveguide is robust against interface fluctuations, which is very useful in practical applications

## Conclusion

In conclusion, based on 2D TLs with lumped elements, we experimentally demonstrate the polarization-dependent selective excitation of guided modes in ENG/MNG waveguides. This PSHE phenomenon is maintained even when the interface has some fluctuations, which offers important applications for spin-directional optical interfaces. Our experimental work further demonstrates the previous theoretical prediction that PSHE is a fundamental phenomenon for surface guided modes.

## Methods

A commercial software package (CST Microwave Studio) is used in designing the samples. The samples are all fabricated on copper-clad 1.6 mm thick FR4 substrates using laser direct structuring technology (LPKF ProtoLaser 200). In the experiment, the signal emission from the port one of vector network analyzer (Agilent PNA Network Analyzer N5222A) and another antenna (i.e., near-filed probe) connecting to the port 2 of analyzer are employed to measure the magnetic fields. A circular probe is vertically placed 1mm above the TLs to measure the signals of magnetic field of the TLs. The field amplitudes are normalized according to their respective maximum amplitude.

# Figure Legends

**Figure 1 | Schematic of the PSHE in an ENG/MNG waveguide.** A circularly polarized source only couples with one guided mode with a specific propagating direction, which is determined by the rotation direction of the source.

**Figure 2 | Effective parameters and the dispersion relationship based on the TLs.** (a) Effective parameters of two kinds of materials in a waveguide (see inset). The parameters can be divided into three regions labeled by three different colors. The permittivity and permeability are shown by the dashed and solid lines, respectively. (b) Dispersion relations of guided modes in ENG/MNG waveguides for TE polarization (the red line) and TM polarization (the blue line), respectively.

**Figure 3 | Transverse spin ($\vec{S}_{surf}$) and the spin-orbit locking in an ENG/MNG waveguide.** For the spin-up ($\vec{S}_{surf} > 0$), the wave-vector or momentum in the MNG (ENG) metamaterial is positive (negative). However, for the spin-down ($\vec{S}_{surf} < 0$), the momentum in the MNG (ENG) metamaterial is negative (positive). The 2D magnetic-field distributions in the $z$ direction at frequency of 2.7 GHz for rotating source with different handedness located in ENG or MNG metamaterial are shown in the four insets.

**Figure 4 | A linearly source couple equally with two guided modes oppositely propagating along the interface.** (a) The sample contains an effective ENG metamaterial loaded with shunted lumped inductors in the upper part and an effective MNG metamaterial loaded with series lumped capacitors in the lower part. A linearly polarized source is loaded near the center of the sample marked by the red point. Inset shows the amplified capacitor and resistor, respectively. The corresponding

circuit model is also given. The simulated (b) and the measured (c) normalized $|H_z|$ patterns of the sample in (a) at 2.7 GHz are given.

**Figure 5 | A circularly polarized source will excite one guided mode with a specific propagating direction**. (a) The sample is similar to that in Fig. 4. However, the source is equivalent to a circularly polarized one which is constituted by four linearly polarized sources (marked by the four red points) with 90 degree phase difference. (b, c) The simulated and the measured normalized $|H_z|$ patterns for the clockwise-rotating source, respectively. (d, e) The source is changed to the anticlockwise-rotating source. (f) The normalized simulated values of $|H_z|$ (the solid lines) along the $x$ direction of ENG/MNG interface. The clockwise-rotating source (CRS) only excites the guided mode propagating from right to left. On the contrary, the anticlockwise-rotating source (ACRS) only excites the guided mode propagating from left to right. (g) The normalized measured values of $|H_z|$ in the same region which are shown by the dashed lines.

**Figure 6 | PSHE in an ENG/MNG waveguide is robust against the interface fluctuations**. (a) Schematic of the PSHE in an imperfect ENG/MNG interface with some bulges. (b, c) The simulated normalized $|H_z|$ patterns for the clockwise-rotating source and the anticlockwise-rotating source, respectively. The sample is the same as that in Fig. 5(a) except that the interfaces are different.

## Acknowledgments

This work is sponsored by the National Key Research Program of China (No. 2016YFA0301101), by the National Natural Science Foundation of China ( No. 11474220, No. 11234010, No. 11264003 and No. 61661007), by Natural Science Foundation of Shanghai (No. 17ZR1443800), and by the Guangxi Natural Science Foundation (No. 2016GXNSFAA380198).

## Author Contributions

Z.W.G. carried out the numerical simulations and experimental measurements; Y.L. conducted the theoretical calculation; H.C., J.R., K.Y., and C.H.X. suggested the approach; H.T.J. supervised the whole work. All authors reviewed the paper.

**Competing financial interests:** The authors declare no competing financial interests.

**Figures**

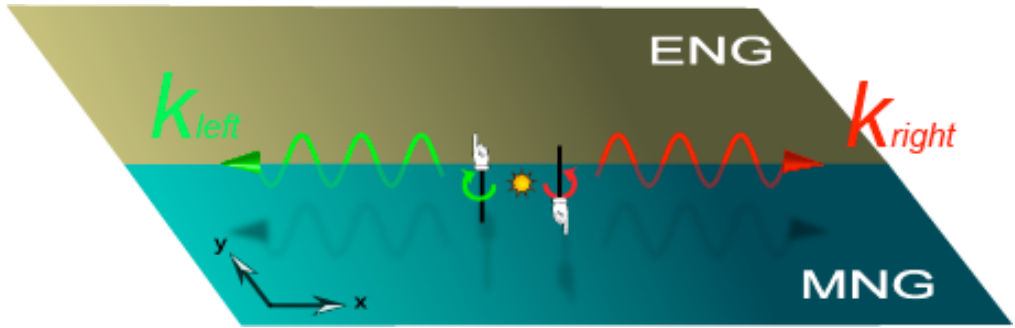

FIG.1

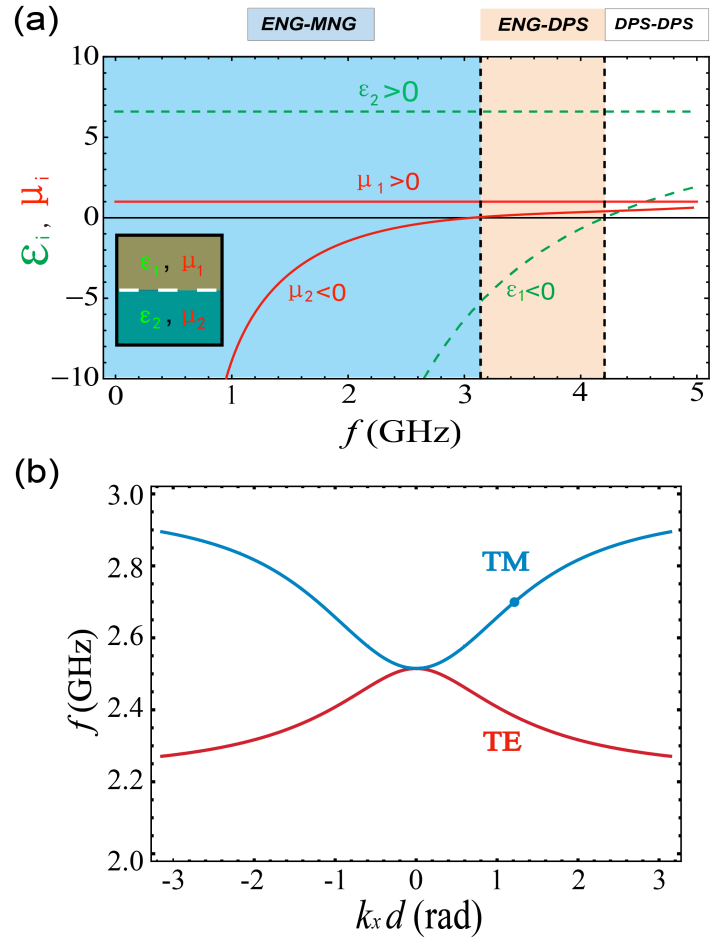

FIG.2

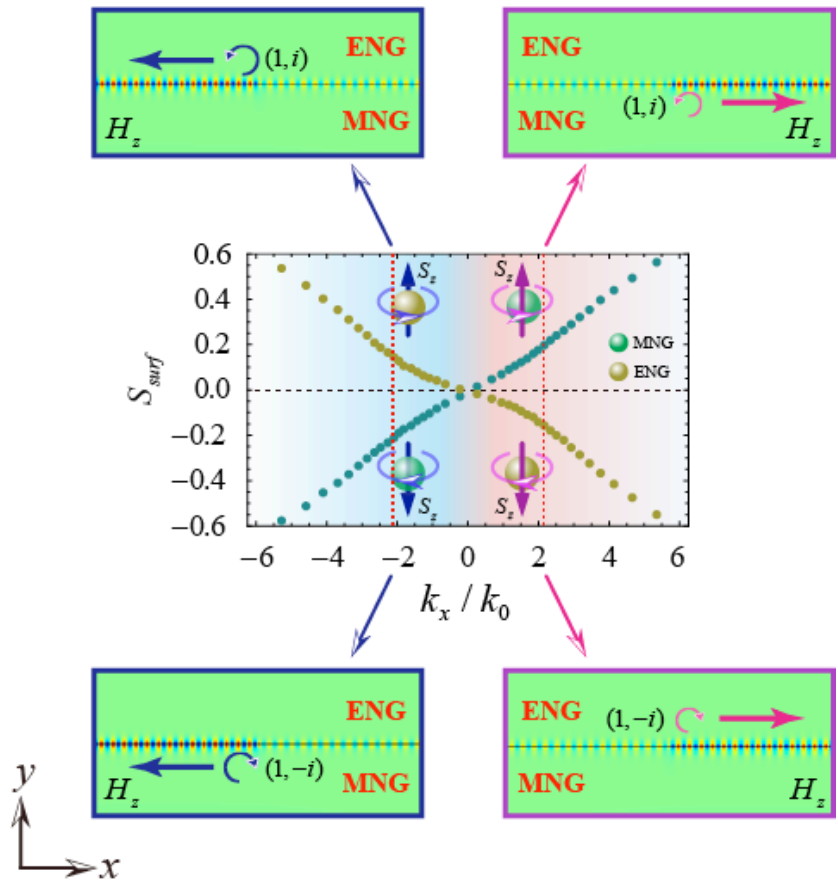

FIG.3

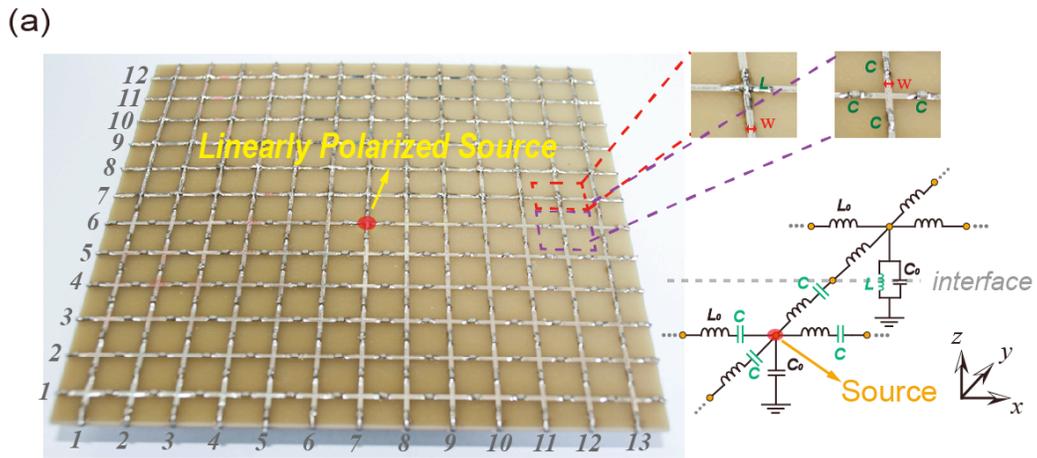

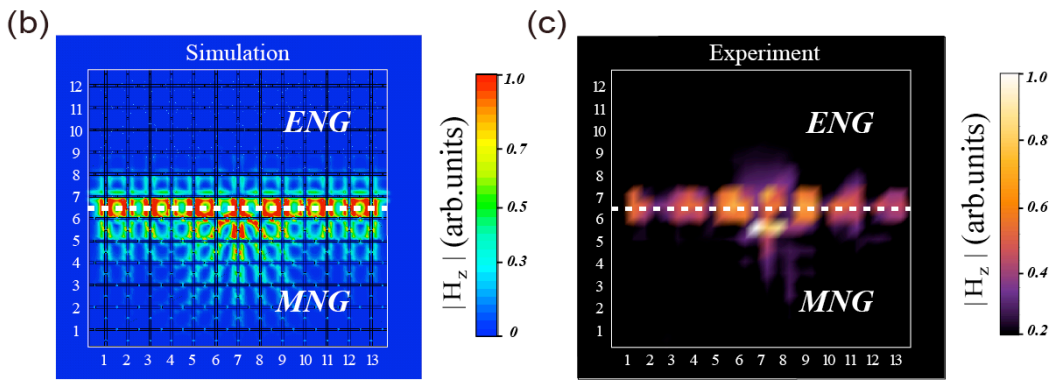

FIG.4

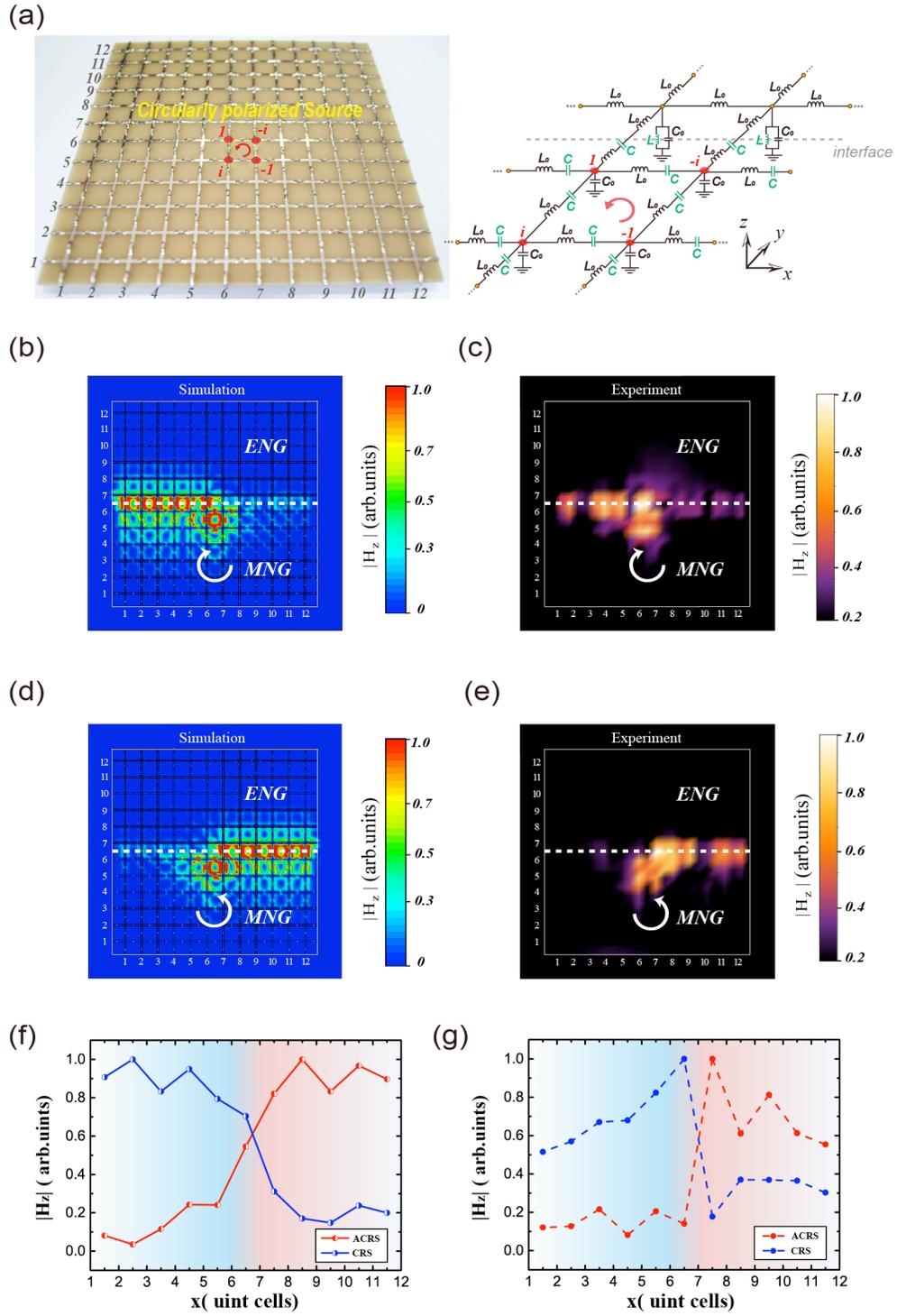

FIG.5

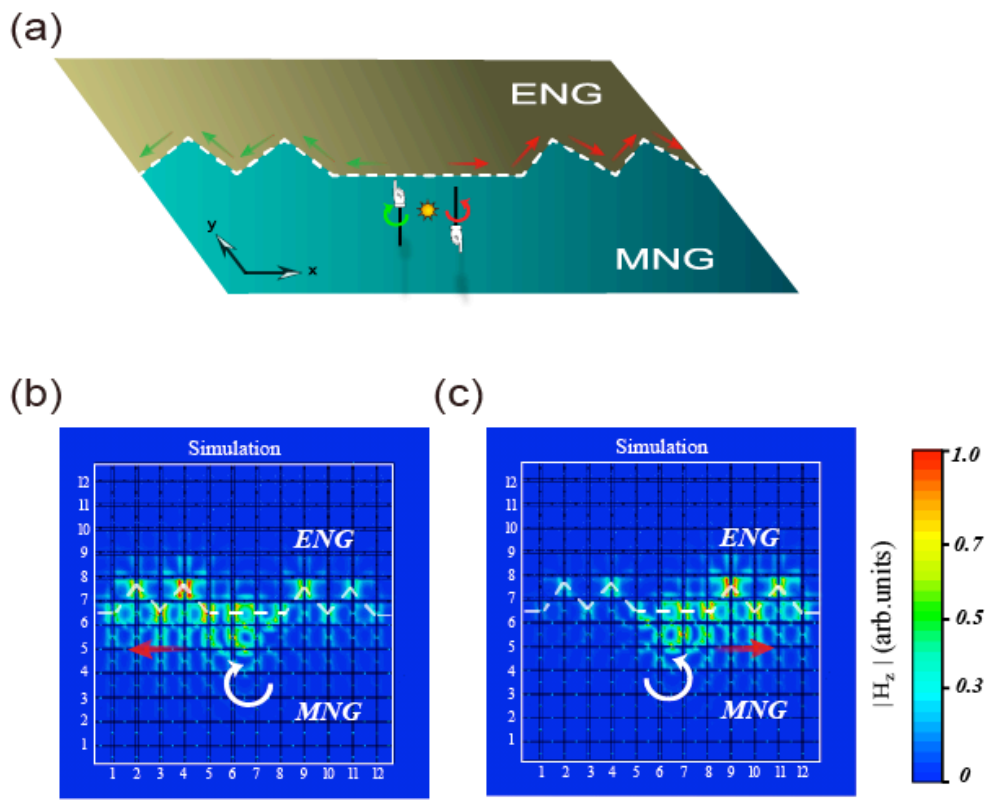

FIG.6